\documentclass[twocolumn,showpacs,preprintnumbers,amsmath,amssymb]{revtex4}

\usepackage{graphicx}
\usepackage{dcolumn}
\usepackage{bm}

\begin{document}

\preprint{APS/123-QED}

\title{Phase transition in a stochastic prime number generator}


\author{Bartolo Luque$^1$, Lucas Lacasa$^{1,*}$ and Octavio Miramontes$^2$}
\affiliation{$^1$Departamento de Matem\'atica Aplicada y Estad\'istica,
ETSI Aeron\'auticos, Universidad Polit\'ecnica de Madrid, Madrid 28040, Spain\\\\
$^2$Departamento de Sistemas Complejos, Instituto de F\'isica, Universidad Nacional Aut\'onoma de M\'exico, 04510 DF, Mexico}%

\date{\today}

\begin{abstract}
We introduce a stochastic algorithm that acts as a prime number
generator. The dynamics of such algorithm gives rise to a continuous
phase transition which separates a phase where the algorithm is able
to reduce a whole set of integers into primes and a phase where the
system reaches a frozen state with low prime density. We present
both numerical simulations and an analytical approach in terms of an
annealed approximation, by means of which the data are collapsed. A
critical slowing down phenomenon is also outlined.
\end{abstract}

\maketitle 

From the celebrated coincidence in $1972$ between H. Montgomery's
work on the statistics of the spacings between zeta zeros and F.
Dyson's analogous work on eigenvalues of random matrices, we have
seen, somewhat unexpectedly, how number theory and physics have
built bridges between each other. These connections range from the
reinterpretation of the Riemann zeta function as a partition
function \cite{julia} or the focus of the Riemann Hypothesis via
quantum chaos \cite{berry}, to multifractality in the distribution
of primes \cite{wolf} or computational phase transitions in the
number partitioning problem \cite{Mertens}, to cite
but a few (see \cite{web} for an extensive bibliography).\\
 Prime numbers are mostly
found using the classical sieve of Eratosthenes and its recent
improvements \cite{factor}. Additionally, several methods able to
generate probable primes have been put forward \cite{factor2}. In
this paper we study a somewhat different algorithm from those
mentioned above, based on an artificial chemistry model introduced
by Dittrich \cite{dittrich2001} that generates primes by means of a
stochastic integer decomposition.\\
Suppose a pool of positive integers $\{2,3,...,M\}$, from which a
subset of $N$ numbers is randomly extracted. Notice that the number
$1$ is ignored and that repetitions are allowed in the subset. Given
two integers $a$ and $b$ (taken from the subset of $N$ numbers), the
reaction rules of the algorithm are:
\begin{itemize}
\item Rule $1$: If $a=b$ then no reaction takes place, and the numbers are not modified.
\item Rule $2$: If the numbers are different, say $a > b$, a reaction will take place only
 if $b$ is a divisor of $a$, i.e. if there exists an integer $c=a/b$.
 Then $a$ is eliminated from the subset and substituted by $c$.
 \item Rule $3$: On the other hand, if $a$ is not divisible by $b$, then no reaction takes place.
\end{itemize}

The stochastic algorithm goes as follows: after choosing a subset of
$N$ numbers from the pool $\{2,3,...,M\}$  two numbers $a$, $b$
belonging to that subset are picked at random; then the reaction
rules are applied to them. We consider $N$ repetitions of this
process as one time step in order to have a parallel updating.
Notice that the positive reactions tend to decompose numbers,
thereby this process when iterated may generate primes. We say that
the system has reached the steady state when no more reactions are
achieved, either because every number has become a prime or because
rule $2$ cannot
be satisfied anymore: the algorithm then stops.\\
In what follows we will firstly present the phase transition that
the system seems to exhibit. Second, we will try to interpret this
phase transition in terms of a dynamical process embedded in a
directed catalytic network, introducing subsequently a proper order
parameter. Some analytical arguments in terms of an annealed
approximation will then be outlined in a third part, where a data
collapse is provided. Finally, a critical slowing down
(easy-hard-easy pattern in the algorithmic phase transitions
language) is pointed out.
\begin{figure}[h]
\centering
\includegraphics[width=0.45\textwidth]{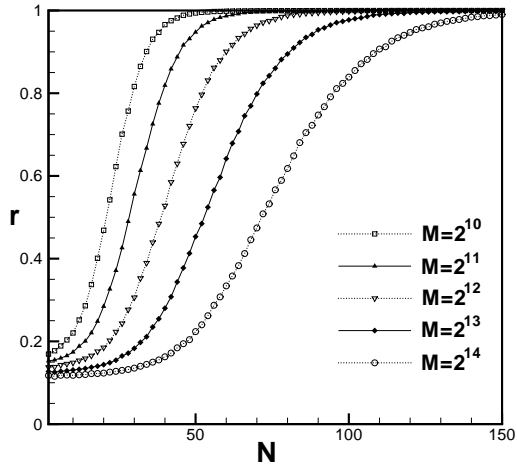}
\caption{The ratio $r$ of prime numbers in the steady state as a
function of $N$ for different pool sizes $M$. Each point is the
average of $2\times10^4$ realizations. For small values of $N$, the
value of $r$ is only related to the amount expected from a random
sample. For large values of $N$, $r$ tends to one: the algorithm is
able to reduce the whole system into primes.} \label{figure1}
\end{figure}

A preliminary indicator of the system's behavior may be the unit
percentage or ratio of primes $r$, that the system reaches in the
steady state. This parameter will characterize the ability of the
algorithm to produce primes, for a given $N$. In figure
\ref{figure1} we plot the behavior of $r$ versus $N$ for several
pool sizes $M$. We clearly see that two separated regimes arise: the
first one is characterized by small ratios (low proportion of primes
in the stationary state) while in the second one every single number
of the system will end up as a prime. The system thus exhibits a
sort of phase transition. Note that $r$ is not a well defined order
parameter since it does not vanish in the disordered phase. This is
due to the fact that following the prime number theorem
\cite{mollin1998}, in a pool of size $M$ there are typically
$M/\ln(M)$ primes. This residual value of $r$ is not related to the
algorithm dynamics. In fact, when $N$ is small the number of
reactions until the system reaches the steady state is quite small.
Therefore, the residual ratio $r\approx1/\ln(M)$ is the relevant
contribution \cite{comentarios}. It comes thus necessary to define
an adequate order parameter that would properly describe the former
phase transition.
\begin{figure}[h]
\centering
\includegraphics[width=0.45\textwidth]{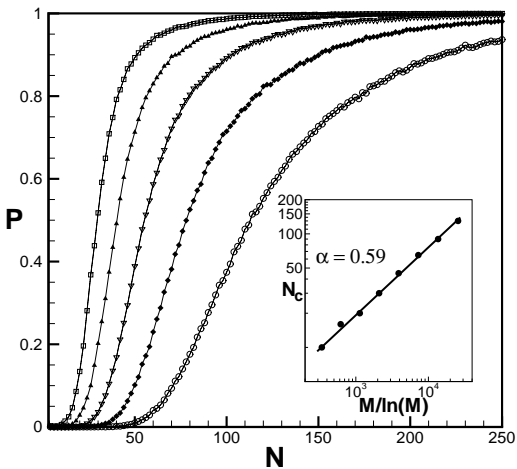}
\caption{Order parameter $P$, defined as the probability that all
numbers are prime in the steady state versus the control parameter
$N$, for the same pool sizes as for figure \ref{figure1}
(simulations are averaged over $2\times10^4$ realizations). Inset:
scaling of $N_c$ versus $M/\ln(M)$, for
 pool sizes $M=2^{10},2^{11},...,2^{17}$.
}
\label{figure2}
\end{figure}

Let us now see how this phase transition can be understood as a
dynamical process embedded in a catalytic network having integer
numbers as the nodes. Consider two numbers of that network, say $a$
and $b$ ($a>b$). These numbers are connected ($a \rightarrow b$) if
they are exactly divisible, that is to say, if $a/b=c$ with c being
an integer. The topology of similar networks has been studied in
\cite{integers1,integers2,integers}, concretely in \cite{integers}
it is shown that this network exhibits scale-free topology
\cite{scalefree}: the degree distribution is $P(k) \sim
k^{-\lambda}$ with $\lambda=2$. In our system, fixing $N$ is
equivalent to selecting a random subset of nodes in this network. If
$a$ and $b$ are selected they may react giving $a/b=c$; in terms of
the network this means that the path between nodes $a$ and $b$ is
traveled thanks to the catalytic presence of $c$. We may say that
our network is indeed a catalytic one \cite{kauffman,kauffman2}
where there are no cycles as attractors but two different stationary
phases: (i) for large values of $N$ all resulting paths sink into
prime numbers, and (ii) if $N$ is small only a few paths are
traveled and no primes are reached. Notice that in this network
representation, primes are the only nodes that have input links but
no output links (by definition, a prime number is only divisible by
the unit and by itself, acting as an absorbing node of the
dynamics). When the temporal evolution of this algorithm is explored
for small values of $N$, we observe that the steady state is reached
very fast. As a consequence, there are only a few traveled paths
over the network and since $N$ is small the probability of catalysis
is small as well, hence the paths ending in prime nodes are not
traveled. We say in this case that the system freezes in a
disordered state. In contrast when $N$ is large enough, many
reactions take place and the network is traveled at large. Under
these circumstances, an arbitrary node may be catalyzed by a large
$N-1$ quantity of numbers, its probability of reaction being high.
Thus, in average all numbers can follow network paths towards the
prime nodes:
we say that the system reaches an ordered state.\\
 In the light of the preceding arguments, it is meaningful to define an order parameter as the probability $P(N,M)$
 that the $N$ numbers extracted from $M$ be primes once the stationary state is reached.
 In figure \ref{figure2} the relation between the order
parameter $P$ and the control parameter $N$ related to the same
simulations that in figure \ref{figure1} is depicted.
 Note that $P$ is now a well defined order parameter, as opposed to
 $r$. In each case, $N_c(M)$ is the critical value separating the
 phases $P=0$ and $P\neq0$. Observe in figure \ref{figure2} that $N_c$ increases with the pool
size $M$.
 In order to describe this size dependence, we need to find some analytical argument by means of which to define
 a system's characteristic size. As we will see below, this one will not be
 $M$ as one would expect.\\
\begin{figure}[h]
\centering
\includegraphics[width=0.45\textwidth]{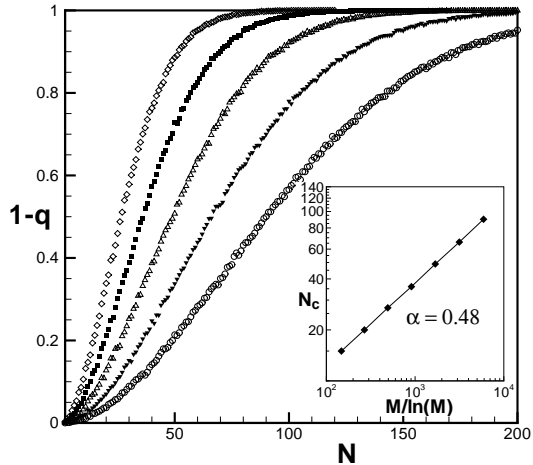}
\caption{ Probability $1-q(N,M)$ that at least two numbers from $N$
be exactly divisible between them, for different values of $M$ (from
left to right: $2^{13},2^{14},2^{15},2^{16}$ and $2^{17}$). Each
point represents the average of $10^5$ realizations. Inset: scaling,
in the annealed system, of $N_c$ (defined such that $q(N_c,M)=0.5$)
versus $M/\ln(M)$ for different values of
$M=2^{11},2^{12},...,2^{17}$.} \label{figure3}
\end{figure}

Note that non-trivial correlations between the values of the $N$
numbers take place at each algorithm's time step. This leads to
highly complex, analytically untractable dynamics. We can try
however an annealed approximation in order to break these
correlations, assuming that at each time step, the $N$ numbers are
randomly generated. In figure \ref{figure3} we depict for different
values of $M$, the resulting simulations of the function $1-q$,
where $q=q(N,M)$ is the probability that no pair of randomly chosen
$N$ numbers from $M$ be divisible between them. Notice that once $M$
is fixed, there is a certain value of $N$ from which the probability
of finding at least one reacting pair is almost $1$. The behavior of
$1-q$ follows qualitatively the behavior of the order parameter $P$.
In fact, this annealed approximation suggests that once $M$ is fixed
in the algorithm, from a certain $N$ we can be sure that at least
one reaction will take place. As long as reactions produce new
numbers while the total number $N$ is conserved, reactions will then
take place until the
system reaches a stationary state of only primes.\\
The probability $p(M)$ of a reaction between two randomly chosen
numbers from the pool $M$, that is to say, the probability that two
numbers from $\{2, 3,..., M\}$ be divisible is:
\begin{equation}
p(M)=\frac{2}{(M-1)^2} \sum_{x=2}^{\lfloor
M/2\rfloor}\bigg\lfloor\frac{M-x}{x}\bigg\rfloor\approx
\frac{2\ln(M)}{M}.
\end{equation}
Obviously, $1-p(M)$ is the probability that two randomly chosen
numbers from $\{2, 3,..., M\}$ are not divisible. From a set
containing $N$ randomly chosen numbers, the $N(N-1)/2$ different
pairs that we can form are not independent, therefore the
probability $q(N,M)$ is not simply $(1-p(M))^{N(N-1)/2}$.
Correlations between pairs can however be taken into account through
the following ansatz:
\begin{equation}
q(N,M)\approx \left(1-\frac{2\ln(M)}{M}\right)^{N^{1/\alpha}},
\end{equation}
where the exponent $\alpha$ characterizes the degree of dependence
between pairs.  For convenience, we assume that the threshold
 $N_c(M)$ in this annealed approximation is the one for which half of the configurations reach an ordered state, that is to say,
 the values for which $q(N_c,M)=0.5$. This procedure is usual for instance in percolation processes,
 since the choice of the percolation threshold, related to the definition of a spanning
cluster, is somewhat arbitrary in finite size
 systems \cite{percolation}.
After some algebra and taking a leading-order approximation, we find
the scaling relationship:
\begin{equation}
N_c\sim\left(\frac{M}{\ln(M)}\right)^{\alpha}.
\label{scaling_teorico}
\end{equation}
In the inset of figure \ref{figure3} we plot in log-log the scaling
between $N_c$ and $M/\ln(M)$ in the annealed system, which follows
equation (\ref{scaling_teorico}) with $\alpha=0.48\pm0.2$ (note that
within the error bar, there is indeed independence between pairs).
The goodness of the
former scaling suggests that the above ansatz is acceptable.\\
\begin{figure}[h]
\centering
\includegraphics[width=0.45\textwidth]{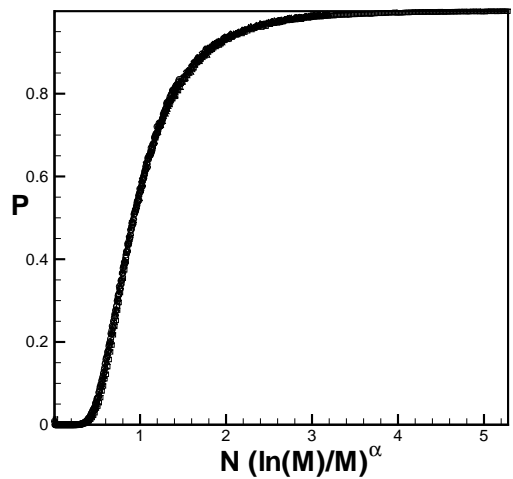}
\caption{Collapse of the order parameter $P$ for different values of
$M=2^{14},2^{15},2^{16},2^{17}$, with $\alpha=0.59$.
} \label{figure4}
\end{figure}
In appearance, in the prime number generator system, the
characteristic size is $M$, however the annealed approximation
suggests that the true characterization is $M/\ln(M)$. From the
point of view of the network this is very reasonable since the
amount of primes that we can reach increases with $M$ in a
non-linear trend, in fact it grows asymptotically as $M/\ln(M)$
\cite{mollin1998}. Coming back to the prime generator system, in
order to prove our foregoing conjecture, in the inset of figure
\ref{figure2} we plot in log-log the values of $N_c$ versus
$M/\ln(M)$. The scaling suggests the same relationship as equation
(\ref{scaling_teorico}) with a scaling exponent
$\alpha=0.59\pm0.05$, that is
to say, we find that the transition point shows critical behavior, as expected.\\
In order to seek consistency, in figure \ref{figure4} we collapse
several curves $P(N,M)$ for different pool sizes $M$. For that task
we apply generic techniques of finite size scaling, where the size
scaling is given by the function $G(M/\ln(M))=(M/\ln(M))^\alpha$,
with $\alpha=0.59$. Note that the data collapse is excellent. This
fact gives credit to the scaling ansatz and provides consistency to
the full development. As long as $N$ is indeed an extensive variable
and in order to find the transition point in the thermodynamic
limit, it is meaningful to define a reduced control parameter
$n=\frac{N}{M/\ln(M)}$, which is now an intensive variable. In the
thermodynamic limit,
we find $n_c=0$.\\
Some other ingredients characterizing the phase transition can be
put forward. First, we may argue that the responsible for the phase
transition is a breaking of symmetry between steady state
distribution of the $N$ integers. As a matter of fact, we can
distinguish a disordered phase ($N<<N_c$) where the steady state
distribution of the $N$ numbers is uniform (each number appears with
the same probability) from an ordered phase ($N>>N_c$) where this
distribution is in turn a power law \cite{comentarios}.\\
Second, in figure \ref{t_sin_colapso} we plot for different pools
the behavior of the characteristic time $\tau$ versus $N$.
$\tau(N,M)$ is defined as the number of time steps per number that
the algorithm needs to perform in order to reach the steady state:
this parameter characterizes the relaxation time of the algorithm.
Note that in each curve $\tau(N,M)$ reaches a peaked maximum in a
neighborhood of $N_c(M)$ (any shift is due to finite size effects).
Moreover, for larger pools this maximum is larger, diverging in the
thermodynamic limit: this behavior is related with a critical
slowing down phenomenon \cite{comentarios}, which in algorithmic
phase transitions is known
as an easy-hard-easy pattern.\\
\begin{figure}[t]
\centering
\includegraphics[width=0.45\textwidth]{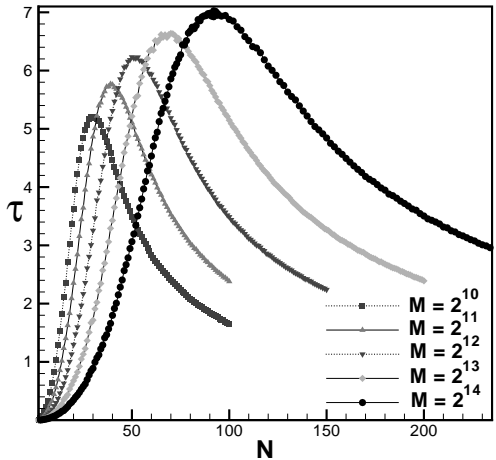}
\caption{Characteristic time $\tau$ as defined in the text versus
$N$, for different pool sizes, from left to right:
$M=2^{10},2^{11},2^{12},2^{13},2^{14}$. Every simulation is averaged
over $2\cdot10^4$ realizations. Note that for each curve,
$\tau(N,M)$ reaches a maximum in a neighborhood $N_c(M)$.}
\label{t_sin_colapso}
\end{figure}

In summary, in this paper we explored a stochastic algorithm that
 works as a prime number generator. Many ingredients suggest the presence of a phase transition in the system.
 This unexpected behavior raises some interesting related questions that will be considered in further work, namely: how the character of
 such transition is related to with the computational complexity of the algorithm \cite{TF}? In which way the algorithm, which produces
 primes by means of stochastic decomposition, is related to the integer factorization problem and cryptography?\\

The authors wish to thank D. Zanette, U. Bastolla, A. Robledo and G.
Garc\'ia-Naumis for reading our manuscript and providing helpful
suggestions, and the Instituto de F\'isica at UNAM for support and
hospitality. The research was supported by UNAM-DGAPA grant
IN-118306 (OM), grants FIS2006-26382-E and FIS2006-08607 (BL and LL)
and the \emph{Tomas Brody Lectureship 2007} awarded to BL.

\small{$^*$\emph{Electronic address}: lucas@dmae.upm.es.}


\begin{thebibliography}{10}
\bibitem{comentarios} Additional data and simulations concerning this phenomenon will be provided in a further work.
\end{thebibliography}
\end{document}